\documentclass[fleqn,twoside]{article}
\usepackage{espcrc2}

\usepackage{psfig}

\title{{\tiny \hfill{FZJ-IKP-TH-2004-7, HISKP-TH-04/10}}\\[0.5em]
Determination of the ${\rm {\bar K^0}d}$ scattering length from the 
reaction ${\rm pp{\to}d{\bar K^0}K^+}$}
\author{A.~Sibirtsev\address[i1]{Institut f\"ur Kernphysik, 
Forschungszentrum J\"ulich, D-52425 J\"ulich, Germany}, 
M.~B\"uscher\addressmark[i1], V.Yu.~Grishina\address[i2]{Institute
for Nuclear Research, 60th October Anniversary Prospect 7A, 
117312 Moscow, Russia},
C.~Hanhart\addressmark[i1],  
L.A.~Kondratyuk\address[i3]{Institute of 
Theoretical and Experimental Physics, 
B. Cheremushkinskaya 25, 117218 Moscow, Russia}, 
S.~Krewald\addressmark[i1], \\
U.-G.~Mei{\ss}ner\addressmark[i1]~\address[i2]
{Universit\"at Bonn, Helmholtz-Institut f\"ur Strahlen und 
Kernphysik (Theorie),  Nu{\ss}allee 14-16, D-53115 Bonn, Germany}}

\begin{document}

\begin{abstract}
The real and imaginary parts of the ${\bar K^0}d$ scattering length  
are extracted from the  ${\bar K^0}d$ mass spectrum obtained 
from the reaction $pp{\to}d{\bar K^0}K^+$  measured recently at 
the Cooler Synchrotron COSY at J\"ulich.
We extract a new limit on the $K^-d$ scattering length, namely
${\rm Im}~a{\le}$1.3~fm and $|{\rm Re}~a|{\le}$1.3~fm.
The limit for the imaginary part of the $K^-d$ scattering length is
supported by  data on the total  $K^-d$ cross sections.
\end{abstract}

\maketitle

During the last two decades the physics of the low-energy ${\bar K}N$ 
and ${\bar K}A$ interactions has gained substantial interest. 
A well-known $K$-matrix analysis~\cite{Martin} of 
the available $K^-N$ data led to the conclusion that the real 
part of the $s$-wave 
$K^-p$ scattering length is repulsive, ${\rm Re}~{a_{K^-p}}{=}-0.7$~fm, 
while the real part of the $K^-n$ scattering length is attractive, 
${\rm Re}~{a_{K^-n}} {=} 0.37$~fm. In a recent KEK experiment the strong 
interaction shift of the kaonic hydrogen atom $1s$ state was found 
to be repulsive~\cite{Ito}, corresponding to a negative 
$K^-p$ scattering length.  The first results for kaonic hydrogen from 
the DEAR experiment also indicate a repulsive 
shift~\cite{Guaraldo1}.
However, at the same time there are no direct
experimental results available for the  $K^-n$ scattering length. 
{}From a theoretical point of view it is natural to expect
that the $K^-N$ interaction, averaged over proton and neutron targets, 
is attractive. One of the fundamental reasons for
this expectation is given by the leading order term in the chiral 
expansion for the $K^-N$ channel which appears to be attractive
(in contrast to the isoscalar pion-nucleon scattering amplitude).
In fact it is possible to have a negative scattering length 
${\rm Re}~{a_{K^-p}}$ for the attractive
${\bar K}N$ interaction if the $\Lambda(1405)$ resonance is a bound state
of $\bar{K}N$ system~\cite{Dalitz,Weise1}. Such a peculiar dynamics of
the elementary ${\bar K}N$ interaction implies  non-trivial properties
of anti-kaons in finite nuclei and dense nuclear matter, including neutron 
stars, see e.g. Refs.~\cite{Lutz,Ramos,Heiselberg,Cieply} 
(and references therein). 
A renewed interest in physics with low--energy kaons was initiated 
by substantial progress in effective low energy hadronic methods, 
in particular by approaches based on  extensions of chiral
perturbation theory, for early reviews see 
Refs.~\cite{Meissner,Ecker,Pich,Bernard}. 
The coupled channel dynamics of the ${\bar K}N$ interaction
based on tree level chiral Lagrangians was developed in order 
to describe low-energy scattering data~\cite{Weise,Oset,Oller} and
giving further support to the description of the $\Lambda (1405)$ as
a meson-baryon bound state (a new twist on this story was given in 
Ref.~\cite{Jido}, where the two-pole structure of this 
state was investigated). 
However, it was shown  recently~\cite{Meissner1} 
that a reliable extraction  of the elementary ${\bar K}N$ scattering length 
from such type of approach requires an  explicit
matching of the amplitudes generated from the coupled channel dynamics
to the ones given from  chiral perturbation theory~\cite{Oller2}. 
This matching can even be done in some unphysical region of the
corresponding Mandelstam plane. If this is not done,
the calculations result in very large $K^-p$ and $K^-n$
scattering lengths, which contradict the experimental results~\cite{Martin}
and might stem from the implicit violation of chiral symmetry.
Therefore, new and more exact experimental results on $K^-p$ and 
$K^-n$ scattering are necessary in order to obtain reliable constraints 
on the $K^-N$ dynamics and to gain a  better understanding of the  
description of SU(3) chiral symmetry breaking. The measurement 
of the $K^-d$ scattering length is one of the main goals of the SIDDHARTA 
experiment at DA$\Phi$NE~\cite{Marton}. In \cite{Meissner1}
the precise relation between the energy-shift in kaonic hydrogen
and the appropriate scattering lengths combination is worked out.

In this paper we show that  constraints on the ${\bar K}d$ 
scattering length can be obtained through an analysis of the 
${\bar K^0}d$ final--state interaction (FSI) in the reaction 
$pp{\to}d\bar{K^0}K^+$ near  threshold measured very recently 
at COSY~\cite{Kleber}.
Theoretical analyses of the reaction $pp{\to}\bar{K^0}dK^+$
near threshold have been performed  in 
Refs.\cite{Grishina2000,Meissner2,Grishina2001,Kudryavtsev,Hanhart,Grishina04}.
As has been stressed in Ref.~\cite{Meissner2}, the  
$NN{\to}d{\bar K}K$ reaction
should be sensitive to the ${\bar K}d$ FSI. Thus one can expect that the
experimental results on the $pp{\to}d\bar{K^0}K^+$ reaction may
provide a new way to extract the $\bar{K^0}d$ scattering length.

In this paper we focus on the potential influence of the $\bar K d$
interaction on the observables for the process $pp{\to}d\bar{K^0}K^+$. 
The effect of the $\bar K K$ $s$--wave interaction can be investigated via the
inclusion of a Flatt\'e distribution for the $a_0 (980)$. 
This had a negligible effect on the shape spectra \cite{Kleber}. 
A study of a possible interplay of the $\bar K d$
and the $\bar K K$ interaction will be presented in a subsequent publication.
However, the data is compatible with a quite weak $\bar K d$ interaction only
and therefore adding the $\bar KK$ interaction is not expected to 
change the picture significantly.

Near threshold the partial wave structure of the final--state for the 
$pp{\to}d\bar{K^0}K^+$ reaction can be described by the superposition 
of two configurations $[(\bar{K^0}K^+)_s d]_P$ and $[(\bar{K^0}K^+)_p d]_S$ 
with the $\bar{K^0}K^+$-system in the $s$- and $p$-wave, respectively.
Correspondingly, the deuteron is in a $p$-wave (or $s$-wave) with respect
to the  $\bar{K^0}K^+$-system in the first (or second) case.  
An overall $s$--wave is forbidden by  selection rules.

Therefore, if we restrict ourselves to these partial waves,
the most general spin--averaged squared matrix element is  given as
\begin{eqnarray}
&&\!\!\!\!\!
\overline{|M({\bf q,k}) |^2}{=}C_0^q q^2{+}C_0^k k^2{+}
C_1 ({\bf{\hat p}}\cdot{\bf k})^2 \nonumber \\
&&\!\!\!
{+}C_2( {\bf \hat p}\cdot{\bf q})^2 
{+}C_3 ({\bf k {\cdot}q})+ C_4({\bf \hat p}\cdot {\bf k})
({\bf \hat p}\cdot {\bf q}) 
\label{C_i}
\end{eqnarray}
where $\hat{\bf p}{=}{\bf p}/|{\bf p}|$ and $\bf{p}$ is the initial
center-of-mass (c.m.)  momentum, $\bf{k}$ is the final c.m. momentum of the
deuteron, and $\bf{q}$ is the relative c.m. momentum of the
$\bar{K^0}K^+$-system.  Furthermore, the six real coefficients $C_i$ can be
expressed through the corresponding partial wave
amplitudes~\cite{Grishina2001,Kudryavtsev}.
 
In Ref.~\cite{Kleber} the parameters $C_i$ where extracted from  the data
at a proton beam energy of $T_p$=2.65~GeV. The
parameters $C_0^q$ and $C_2$ account for the contributions from the $K\bar K$
$p$-wave, $C_0^k$ and $C_1$ from the $K\bar K$ $s$-wave and $C_3$ and $C_4$
stem from $s$-$p$ interference.

\begin{figure}[t]
\vspace*{-5.mm}
\hspace*{-2.mm}\psfig{file=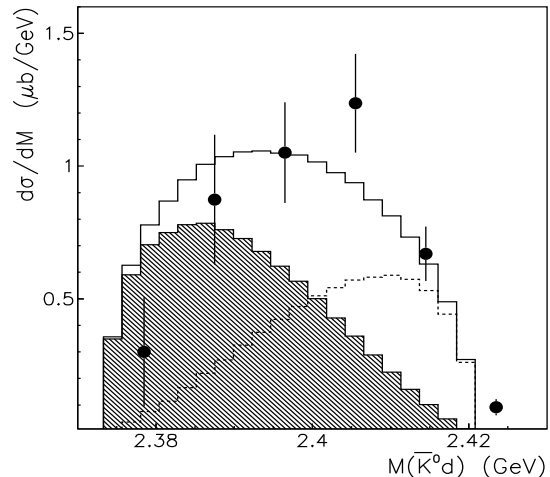,height=7.2cm,width=8.cm}
\vspace*{-17mm}
\caption[]{${\bar K^0}d$ mass spectra from the reaction
$pp{\to}d{\bar K^0}K^+$ 
at $T_p{=}$2.65~GeV, i.e. an excess energy of 46 MeV. 
The experimental results are 
taken from Ref.~\cite{Kleber}. The experimental mass resolution is
$\simeq$3 MeV. The hatched histogram
indicates the ${\bar K^0}d$  $s$-wave contribution, 
dashed- $p$-wave and solid is their sum. }
\label{kdfsi1}
\vspace*{-5.9mm}
\end{figure}

To analyze the ${\bar K^0}d$ FSI in the $s$-wave it is necessary to isolate
the ${\bar K^0}d$ s-wave contribution from the $pp{\to}d{\bar K^0}K^+$
reaction. Therefore we should express the matrix element in terms of the
partial amplitudes in a basis that is different to the one given above, namely
in terms of $[(\bar{K^0}d)_s K^+]_P$ and $[(\bar{K^0}d)_p K^+]_S$ states,
where the $\bar{K^0}d$-system is in the $s$- and $p$-wave, respectively.
As was proposed in Ref.~\cite{Hanhart} the vectors of Eq. (\ref{C_i}) 
can be expressed in terms of the c.m. momentum of the $K^+$, ${\bf P}$, 
and the relative momentum of the $\bar K^0d$ system, ${\bf Q}$, as 
\begin{eqnarray}
{\bf q}{=} {\bf Q} {-} \alpha {\bf P} \, , \quad {\bf k}
{=}\frac{1}{2}\bigl((2-\alpha) {\bf P} {+} {\bf Q}\bigr) \ ,
\end{eqnarray}
where $\alpha{=} m_d/(m_d{+}m_{\bar K})$. The squared amplitude
expressed in the new frame reveals the same structure as
Eq. (\ref{C_i}), namely:
\begin{eqnarray} \nonumber
&&\!\!\!\!\!
\overline{|M({\bf Q,P}) |^2}{=}B_0^{Q}{Q}^2{+}B_0^{P}{P}^2
{+}B_1({\bf  P\cdot \hat p})^2 \\&&\!\!\!\!\!\! {+}B_2({\bf  Q\cdot \hat p})^2
{+}B_3({\bf  P\cdot Q}){+}B_4({\bf  P\cdot \hat p})({\bf  Q\cdot \hat p}) \ ,
\label{mformnew}
\end{eqnarray}
where the $B_i$ coefficients can be expressed in terms of the $C_i$
from Eq. (\ref{C_i}) and
\begin{eqnarray}
\nonumber &&\!\!\!\!\!
B_0^{P}{=}\frac{(2-\alpha)^2}{4}C_0^{k}{+}\alpha^2C_0^{q}{-}
\frac{\alpha(2-\alpha)}{2}\frac{1}{2}C_3
\ ,\\
&&\!\!\!\!\!
B_1{=}\frac{(2-\alpha)^2}{4}C_1^{k}{+}\alpha^2C_1^{q}{-}
\frac{\alpha(2-\alpha)}{2}\frac{1}{2}C_4\ . 
\end{eqnarray}

Using the results of the fit for $C_i$ from Ref.~\cite{Kleber} we obtain the
following values for the coefficients $B_i$: $B_0^Q$=0.81, $B_0^P$=0.705,
$B_1$=-0.267, $B_2$=-0.267, $B_2$=-1.45, $B_3$=1.41. 
It follows from this that the ${\bar K}d$ s-wave contributes 57\% 
to the total cross section.  Fig.~\ref{kdfsi1} shows the ${\bar K^0}d$ 
mass spectra from the $pp{\to}d{\bar K^0}K^+$ reaction at
the beam energy $T_p{=}$2.65~GeV~\cite{Kleber}.
The histograms show our calculations with the parameters $B_i$ given above.
The hatched histogram shows the ${\bar K^0}d$ $s$-wave contribution, the
dashed line illustrates the $p$-wave contribution, while the solid histogram
shows the result of our full calculation.  We recall that in 
Ref.~\cite{Kleber} the parameters $C_i$ were obtained from a joint fit
to ${\bar K^0}K^+$ and ${\bar K^0}d$ mass spectra as well as 
$\cos({\bf pk})$, $\cos({\bf pq})$ and $\cos({\bf kq})$ angular distributions.

Since we isolate the $s$-wave contribution from the  ${\bar K^0}d$ 
mass spectrum it is now possible  to study the interaction between
the final ${\bar K^0}$-meson and the deuteron. 
Following the standard Watson and Migdal 
theorem~\cite{Watson,Migdal,Goldberger} we include 
the FSI by multiplying the $B_0^P P^2$ and 
$B_1(\bf{\hat p}\cdot{\bf P})^2$ terms by the enhancement
factor $|1{-}iQa|^{-2}$, where $a$ is the complex scattering 
length. After that correction we refit the experimental
${\bar K^0}d$ invariant mass distribution with two free parameters,
namely the real and imaginary parts of the ${\bar K^0}d$
scattering length, while keeping the ${\bar K^0}d$ $s$-wave distribution
fixed. We also checked that the influence of this additional
energy dependence does not  significantly change the other observables given
in Ref.~\cite{Kleber} that went into the fit of the $C_i$ parameters.

\begin{figure}[t]
\vspace*{-4mm}
\hspace*{-3mm}\psfig{file=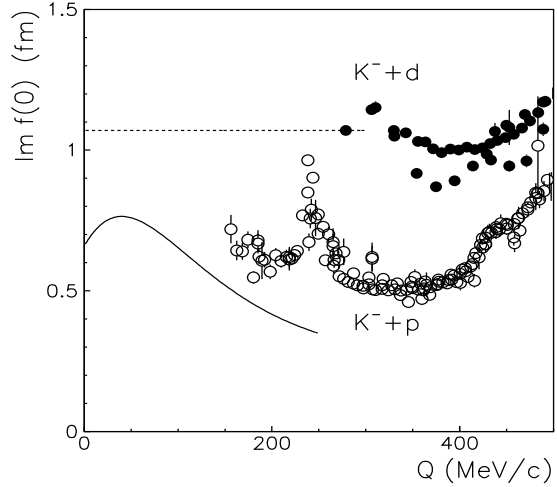,height=7.2cm,width=8.cm}
\vspace*{-16mm}
\caption[]{Imaginary part of the $K^-d$ and the $K^-p$ forward scattering 
amplitudes, respectively, as a function of the cms momentum $Q$. 
The data were obtained 
from total cross sections using the optical theorem. The solid line shows 
the $K$-matrix solution  for $K^-p$ $s$-wave amplitude given by 
Martin~\cite{Martin}. The dashed line shows our extrapolation for    
$K^-d$ amplitude.}
\label{kdfsi2}
\vspace*{-5mm}
\end{figure}

Before performing the fit, 
we have to specify the boundary conditions for the
${\bar K^0}d$ scattering length $a$. While there are no experimental
constraints on the real part ${\rm Re}~{a}$ of the  ${\bar K^0}d$ scattering 
length, it is clear that ${\rm Im}~{a}$ must be positive because of unitarity.
Furthermore, a lower bound on the imaginary part can be deduced from 
the experimental  data on the $K^-d$ total cross section $\sigma_{\rm tot}$
using the
optical theorem for the forward scattering amplitude
\begin{eqnarray}
{\rm Im}~f(0){=}\frac{Q \,  \sigma_{\rm tot}}{4\pi},\,\,\,\,\,\, \,\,\,
a{=}f(0)\left|{_{Q{\to}0}}\right. .
\end{eqnarray}
Fig.~\ref{kdfsi2} shows the imaginary part of the 
$K^-d$ forward scattering amplitude as a function of the c.m. momentum $Q$ 
deduced from the data on the $K^-d$ total cross section (solid circles). 
The extrapolation 
below 300 MeV/c by a straight dashed line gives
${\rm Im}~{a}{\simeq}1.1$~fm. However, one might argue whether this 
momentum independent extrapolation can be considered as being  realistic, 
since the contribution from the $\Lambda{(1405)}$ and $\Sigma{(1385)}$ 
resonances might be quite strong at small $Q$. 

\begin{figure}[t]
\vspace*{-4mm}
\hspace*{-4mm}\psfig{file=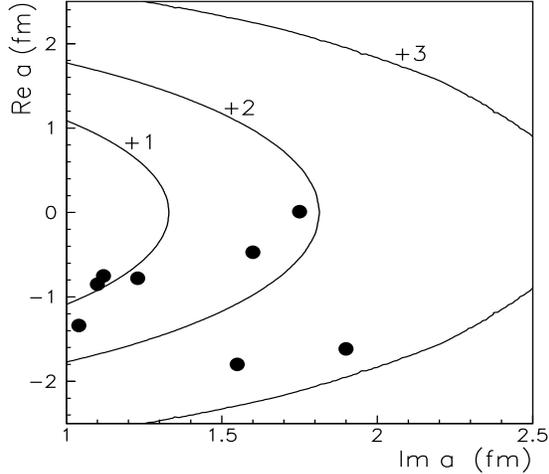,height=7.2cm,width=8.cm}
\vspace*{-16mm}
\caption[]{Real versus imaginary part of the ${\bar K^0}d$ scattering 
length. The solid contour lines show the results of our fit 
to the $pp{\to}d{\bar K^0}K^+$ data~\cite{Kleber} for $\chi^2{+}1$,
$\chi^2{+}2$ and $\chi^2{+}3$. The solid circles show the 
results from  model calculations collected in Table~1.}
\label{kdfsi3}
\vspace*{-6mm}
\end{figure}

To investigate that problem we show by the open circles in Fig.~\ref{kdfsi2}
the imaginary part of the $K^-p$ forward scattering amplitude. 
The extrapolation below 200 MeV/c by the straight line gives for the 
imaginary part of $K^-p$ scattering length $\simeq$0.7~fm. 
This bound for the imaginary part of the $K^-p$ scattering length agrees
 with the $K$-matrix solution for the $K^-p$ $s$-wave 
amplitude found by Martin~\cite{Martin}. Note that the $K$-matrix 
solution includes the $\Lambda{(1405)}$ and $\Sigma{(1385)}$ resonances
and the momentum dependence of $f(0)$ obtained from the $K$-matrix is shown 
by the solid line in Fig.~\ref{kdfsi2}. Obviously the $K$-matrix 
result  underestimates the data on the imaginary part of the $K^-p$
forward scattering amplitude at $Q{\ge}160$~MeV/c, since this
solution does not include all channels that are open at high momenta.
Based on these arguments we 
finally conclude that a reasonable estimate of the lower limit for  
the imaginary part of ${\bar K^0}d$ scattering length is about 1~fm.
Therefore, we fit the experimental results~\cite{Kleber} 
for the ${\bar K^0}d$ invariant mass spectrum using the lower bound
${\rm Im}~a{\geq}1$~fm.

\begin{table}[t]
\caption{The $K^-d$ scattering lengths predicted by different 
calculations with various elementary $K^-p$ and $K^-n$ scattering
lengths. Here, FCA denotes the fixed center approximation, while 
FE stands for the calculations by Faddeev equations. All values are
in fm.}
\begin{tabular}{|l|c|c|}
\hline 
$a(K^- N)$  & $K^-d$   & Ref. \\
\hline \hline
$K^-p = -0.66+i0.67$ & FCA&     \\
$K^-n = 0.264+i0.57$ & $-0.78+i1.23$ & \cite{Grishina04} \\
\hline 
$K^-p = -0.70+i0.71$ & FE&     \\
$K^-n = 0.28+i0.67$ & $-1.34+i1.04$ & \cite{Torres} \\
\hline 
$K^-p = -0.66+i0.67$ & FCA &     \\
$K^-n = 0.26+i0.57$ & $-0.75+i1.12$ & \cite{Deloff} \\
\hline 
$K^-p = -0.66+i0.67$ & FE &     \\
$K^-n = 0.26+i0.57$ & $-0.85+i1.10$ & \cite{Deloff} \\
\hline 
$K^-p = -0.045+i0.835$ & FCA &     \\
$K^-n =0.94+i0.72$ & $-0.01+i1.75$ & \cite{Deloff} \\
\hline 
$K^-p = -0.045+i0.835$ & FE &     \\
$K^-n = 0.94+i0.72$ & $-0.47+i1.60$ & \cite{Deloff} \\
\hline 
$K^-p = -0.789+i0.929$ & FCA &     \\
$K^-n = 0.574+i0.619$ & $-1.615+i1.909$ & \cite{Kamalov} \\
\hline 
$K^-p = -1.01+i0.95$ & FE&     \\
$K^-n =0.54+i0.53$ & $-1.92+i1.58$ & \cite{Bahaoui} \\
\hline 
\end{tabular}
\label{tab1}
\vspace*{-4mm}
\end{table}

The results of our fit are shown in Fig.~\ref{kdfsi3}. With the lower limit
${\rm Im}~a{\geq}1$~fm we obtained the total $\chi^2 = 9.6$. The solid lines
in  Fig.~\ref{kdfsi3} indicate the $\chi^2$+1, $\chi^2$+2 and $\chi^2$+3
contour lines. Furthermore,  the solid circles show different results 
for $K^-d$ scattering length from the calculations
collected in Table~1. It is important to remark that the prediction from
Ref.~\cite{Torres} was based on a combined analysis of the 
experimental results on $K^-d{\to}N\Lambda\pi$ and  
$K^-d{\to}N\Sigma\pi$ relative rates and spectra and is quite
close to our solution.

Our analysis of the ${\bar K^0}d$ mass spectra for the
$pp{\to}d{\bar K^0}K^+$ reaction allows one 
to accept some predictions~\cite{Torres,Deloff,Grishina04}
for the $K^-d$ scattering length  within the range 
${\rm Im}~a{\le}$1.3~fm and $|{\rm Re}~a|{\le}{-}$1.3~fm. The limit for 
the imaginary part of the $K^-d$ scattering length is also strongly supported
by the data on the total $K^-d$ cross section shown in Fig.~\ref{kdfsi2}.
Note that the model results listed in Table~1 have been calculated 
with different input parameters for the elementary $K^-p$ and $K^-n$ 
scattering lengths, also given in  Table~1. The different
input as an elementary $K^-p$ and $K^-n$ scattering lengths  largely 
explains the variations in the final results for the 
$K^-d$ scattering length.

As a next step the elementary  $K^-p$ and $K^-n$ 
scattering lengths might be extracted from our results obtained from the 
$pp{\to}d{\bar K^0}K^+$ data applying established few-body techniques.
In addition the possible interplay of the $\bar KK$ interaction and the $\bar
K d$ interaction, also the influence of a non--vanishing 
$\bar K d$ effective range should be studied. 
 
\medskip

\noindent {\bf Acknowledgements}\\
L.A.K. is grateful to the Institut f\"ur Kernphysik (Theory)
at the Forschungszentrum
J\"ulich for providing hospitality during the course of carrying out 
this work. This work is partially supported by the Russian Fund for 
Basic Research (grants 02-02-1673 and 03-02-04025).

\end{document}